# CuFeAs: A New Member in the 111-Family of Iron-Pnictides

**Gohil S Thakur[1], Zeba Haque[1], L C Gupta[1†] and A K Ganguli[1,2]***

[1]*Department of Chemistry, Indian Institute of Technology, New Delhi 110016, India*

[2]*Institute of Nano Science & Technology, Habitat Centre, Mohali, 160062 India*

We have synthesized CuFeAs, a new iron-pnictide compound with a layered tetragonal $Cu_2Sb$ type structure (space group $P4/nmm$: $a = b$ = 3.7442(2) Å and $c$ = 5.8925(4) Å) that is identical to that of 111-type iron-based superconductors. Our measurements suggest that in low applied magnetic field it undergoes an antiferromagnetic transition below $T_N$ ~ 9 K. When compared with the ground state of CuFeSb, recently reported 111-type ferromagnetic material ($T_C$ ~ 375 K), it has important implication with regard to the nature of Fe-Fe magnetic interaction in Fe-As materials. CuFeAs does not exhibit superconductivity down to 2 K.

**1. Introduction:**

$Cu_2Sb$-type compounds have been known for a long time and their crystal chemistry has been well established in literature.[1] $Cu_2Sb$ crystallizes in a tetragonal structure ($P4/nmm$ space group), with a unit cell containing 4 Cu-atoms and 2 Sb-atoms ($Z$ = 2). The Cu-atoms occupy two nonequivalent crystallographic sites 2(a) (Wyckoff notation, with $4m2$ site-symmetry) designated as Cu(I) and sites 2(c) (with $4mm$ site symmetry) designated as Cu(II). The 2 Sb-atoms also occupy sites 2(c). The two 2(c) sites occupied by Cu(II)-atoms and Sb atoms differ only in the value of 'z' coordinate ($z_1$ ~0.3 and $z_2$ ~ 0.7 respectively) [1]. While the first report on magnetic properties of $Cu_2Sb$ refers to it as antiferromagnetic ($T_N$ ~373 K);[2] later on, using solid state NMR experiments, it was shown to be non-magnetic.[3] Even superconductivity was reported at very low temperature ($T_c$ ~ 0.085 K).[4] Many of the $M_2Pn$ (M = transition element) compounds of $Cu_2Sb$ structure are known to exhibit remarkable magnetic properties. $Mn_2Sb$ is known to order ferrimagnetically well above room temperature ($T_C$ ~ 550K).[5] Some $M_2As$ compounds (M = Mn, Cr and Fe) are known to be antiferromagnetic above room temperature ($T_N$ = 573, 393 and 353 K respectively).[6] Only very recently a new phase CuFeSb has been reported[7] to be ferromagnetic with Curie temperature ($T_C$) ~ 375K. This material has a large $Z_{Sb}$ (Sb height above the Fe-plane) value (~1.84Å). Superconductivity has been observed in several Fe-containing ternary '111' materials AFePn (LiFeP, LiFeAs and NaFeAs)[8-12] that belong to this structure and wherein Li(Na)-atoms occupy the sites 2(c), Fe-atoms occupy the sites 2(a) and the Pn-atoms (P, As) occupy the sites 2(c). Discovery of superconductivity, for example, in LiFeAs ($T_c$ ~16 K)[9] has led to the rapid growth of interest in the 111 materials. In our programme on identifying new Fe-containing superconducting materials, we have synthesized a new 111 material CuFeAs which is different from LiFeAs or NaFeAs in that the 2(c) site are occupied NOT by an alkali metal atom but by Cu, a transition metal atom. With this successful synthesis, a new route of doping Fe-based materials has become available. In this short communication, we report the synthesis and characterization of CuFeAs.

## 2. Experimental

Polycrystalline CuFeAs was synthesized using the following procedure: Stoichiometric proportions of $Cu_2As$ and $Fe_2As$ were heated at 700°C for 24 hrs in evacuated quartz ampoules ($10^{-5}$ Torr) to obtain CuFeAs. $Cu_2As$ and $Fe_2As$ were synthesized by reacting metal and arsenic powders at 800°C for 24 hours in evacuated silica tubes. The resulting binary compounds were thoroughly homogenized using an agate mortar-pestle, pelletized under pressure of 6 tonnes. The polycrystalline product was reground and sintered at 700°C for 4 days in evacuated quartz ampoule. The resulting pellet had a grayish metallic lustre. All the manipulations except pelletizing and sealing were performed inside an argon filled glove box under controlled moisture and oxygen level (< 0.1 ppm each). Many attempts were made to synthesize CuFeAs such as heating the stoichiometric proportions of elements Cu, Fe and As or Cu and FeAs (pre-synthesized) at different temperatures ranging from 680°C to 1000°C to reduce the impurity phases. However in most of our attempts $Cu_3As$ along with main CuFeAs phase invariably appeared as a secondary phase (~15-20%). Our best sample that we used for all our studies reported here has impurity $Cu_3As$ (*) to the extent of ~10% and FeAs ~1%. Powder X-ray diffraction (PXRD) studies were carried out using laboratory X-ray diffractometer (Bruker D8 advance) using CuKα radiation. Structural refinement on powder x-ray diffraction data was carried using Rietveld method with TOPAS package[13]. The impurities were also included in the refinement. Variable temperature magnetization measurements were performed using a commercial SQUID magnetometer (Quantum Design, USA). Resistivity measurements were carried out using conventional four probe method in the temperature range 2-300K.

## 3. Results and discussion

### 3.1 Crystal structure

Powder x-ray diffraction study on our best sample shows that its majority of the sample belongs to the tetragonal $Cu_2Sb$-type phase with small amount of impurity phases, ~ 10 % of $Cu_3As$ (hexagonal) and a tiny amount of impurity FeAs (orthorhombic). Structural refinement using the Rietveld method based on laboratory powder x-ray data is presented in figure 1. The refinement yielded reasonably good reliability factors ($R_{wp}$ 1.93 % and $\chi^2$ = 1.73 %). The crystallographic parameters are tabulated in Table 1. Crystal structure of CuFeAs (see figure 2) consists of α-PbO type layers of edge-sharing $FeAs_{4/4}$ tetrahedra inter-

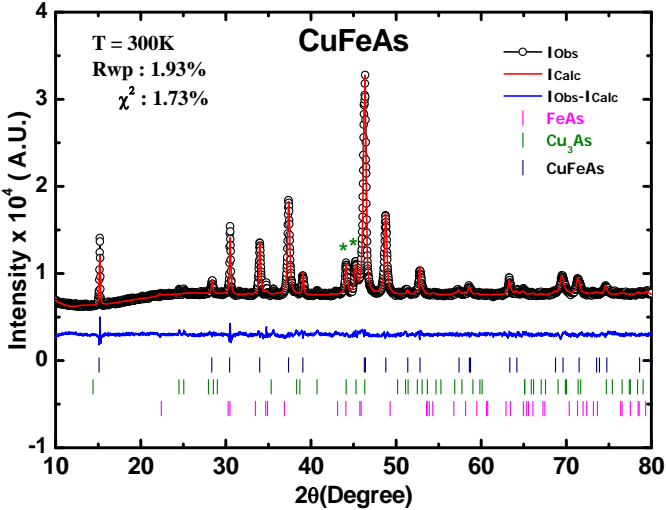

Fig. 1. Powder x-ray diffraction pattern of CuFeAs (black circle) and result of Rietveld refinement (red line). Blue line indicates difference in the experimental and the fitting curve. Vertical bars indicate the Braggs reflections for CuFeAs/impurity phases. **(Color online)**

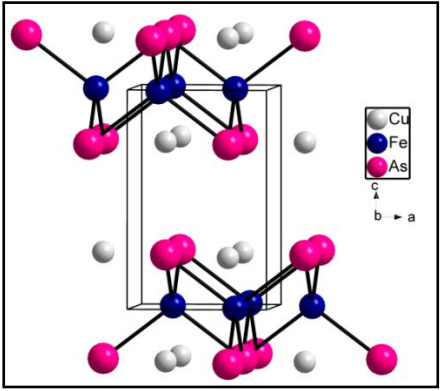

Fig. 2. Crystal structure of CuFeAs. **(Color online)**

| TABLE I. Rietveld refined structural parameters for CuFeAs at 300K. The number in the parenthesis indicates the ESD and the number in the square bracket denotes the multiplicity of the equivalent bond lengths. | | | | | | | |
|---|---|---|---|---|---|---|---|
| Space group $P4/nmm$; $a = b$ = 3.7442(2) Å and $c$ = 5.8925(4) Å: $R_{wp}$ 1.93 % and $\chi^2$ = 1.73 % | | | | | | | |
| | | $x$ | $y$ | $z$ | Occupancy | $B$ (Å$^2$) | Wyckoff positions |
| **Atomic coordinates** | Cu | 0.25 | 0.25 | 0.705(7) | 0.995(2) | 3.3 (2) | 2c |
| | Fe | 0.75 | 0.25 | 0 | 1.030(8) | 0.1 (1) | 2a |
| | As | 0.25 | 0.25 | 0.295(7) | 1.013(5) | 1.0 (2) | 2c |
| **Bond lengths (Å)** | | **Fe-As [4]** 2.555(1) | **Fe-Fe [4]** 2.648(1) | **Cu-As [4]** 2.648(3) | **Cu-As [1]** 2.415(2) | | |
| **Bond angles (°)** As-Fe-As | | $\alpha$ = 94.23 | $\beta$ = 117.59 | | | | |

-spersed with Cu atoms at interstitial sites of As layers, as in LiFeAs (111 type) and similar to $AFe_2As_2$ (122), LnOFeAs (1111) iron pnictide and FeSe/Te (11 type).[3,14-18]

Lattice constants of CuFeAs [$a = b = 3.7442(2)$ Å and $c = 5.8925(4)$ Å] at 300K are smaller than those of CuFeSb ($a = 3.2616$ Å and $c = 6.2515$ Å)[7] due to the smaller size of As atom compared to that of Sb in similar coordination and oxidation state. Despite the ionic sizes of Cu and Li in +1 oxidation state and in similar coordination are almost same, the **c**-lattice constant of CuFeAs is strikingly and unexpectedly smaller than that of LiFeAs ($c = 6.35679$ Å)[9]. The As-Fe-As bond angles in CuFeAs ($\alpha = 94.23°$ and $\beta = 117.59°$) are far from the ideal tetrahedral value (109.5°) and are markedly different from those found in the other families of iron pnictide superconductors.[18] They are, however, similar to those of CuFeSb.[7] The As-height from Fe-Fe plane was calculated to be ~1.74 Å which is significantly larger than in the other iron pnictide families; $Z_{As}$ ~ 1.51 Å for LiFeAs[9], 1.31 Å for LaOFeAs[14], 1.35 Å for $BaFe_2As_2$[15] and $Z_{Se}$ ~1.47 Å for $Fe_{1.01}Se$.[17] It is, however, smaller than $Z_{Sb}$ (1.84 Å) in CuFeSb.[7] The Fe-As distances in CuFeAs (2.555 Å [x4]) are larger than the Fe-As distances in LiFeAs (2.416 Å [x4] at 295 K).[8]

The Cu-As distances in CuFeAs (Cu-As distances are 2.648 Å [x 1] and 2.415 Å [x4]) are smaller than the Li-As distances in LiFeAs (Li–As distances are 2.647 Å [x 1] and 2.759 Å [x 4] at 295 K)[9] owing to the smaller lattice constants and therefore the value of $\alpha$ is smaller and that of $\beta$ is higher. These differences in the structural parameters in CuFeAs and AFeAs may possibly be due to the difference in the bonding characteristics of alkali metal (Li and Na) and Cu with the anion (As and Sb).

## 3.2 Magnetic studies

We have carried out magnetic measurements on CuFeAs from 100 K to 2 K in an applied magnetic field range of 20 - 1000 Oe both in field cooled and zero field condition (Fig. 3.). No diamagnetic response was observed down to 2K indicating absence of superconductivity above 2 K. A prominent cusp like feature at ~ 9 K in FC condition indicates a long range AFM ordering (figure 3 a). In the sample there are two impurity phases namely $Cu_3As$ and FeAs. Since the former is nonmagnetic[19] and the latter orders around 71 K[20] (well above the anomaly temperature observed in our magnetic measurement) it is clear that the peak ~ 9 K is due to AFM ordering of Fe atoms of CuFeAs phase. This cusp like feature is evident in the applied field of 200 Oe also but not as prominent as in the lower field (30 Oe). At higher fields the sample tends to become ferromagnetic (> 500 Oe) which suggests a low flipping field. The M-H curve traced upto 200 Oe (see fig. 4) shows an almost linear behavior which also supports our statement that CuFeAs is antiferromagnet. However at higher applied field the M-H curves attains an s-shape which indicates saturation of the moments (see inset of fig. 4). At zero field there is present a spontaneous magnetization of a magnitude which is much smaller than the magnetization measured at 200 Oe. This we think is due to the possibility that the sample may have a small ferromagnetic component which is not detected in x-ray diffraction studies. Experiments like neutron diffraction would be helpful in sorting out the actual magnetic order in the sample.

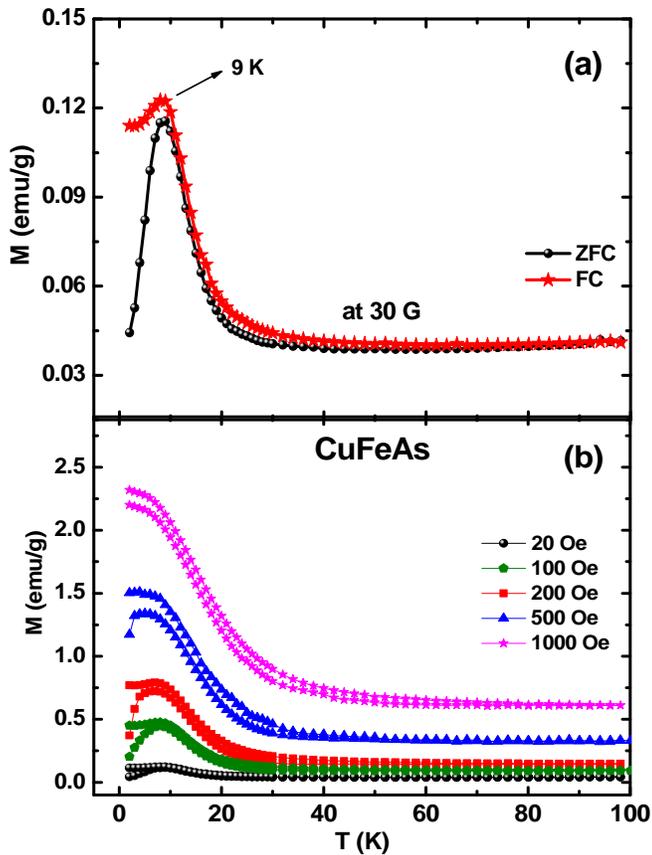

Fig. 3. (a) Magnetisation M(*T*) as a function of temperature in FC and ZFC processes for CuFeAs measured at 30 Oe There is 0a well defined maximum at $T_N \sim 9$ K. (b) M(*T*) at different fields ( 20 - 1000 Oe). See text. **(Color online)**

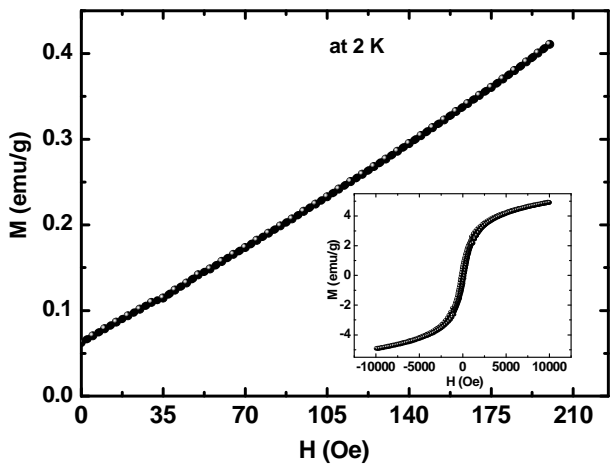

Fig. 4. Field dependent magnetization plot for CuFeAs at 2 K upto 200 Oe. Inset shows M-H loop upto 1 T field. See text.

*3.3 Resistivity studies*

Resistivity measurement down to 2 K (Fig. 5) shows the metallic behavior of the sample with resistivity decreasing monotonously from RT ($\rho = 0.32$ m$\Omega$-cm) to 40K ($\rho = 0.10$ m$\Omega$-cm). Below 40 K, resistivity curve starts flattening may be due to slowing down of spin fluctuation as the sample approaches magnetic order. There is a small change in the slope ~ 220 K also which we suspect is due to the presence of impurity $Cu_3As$.[19]

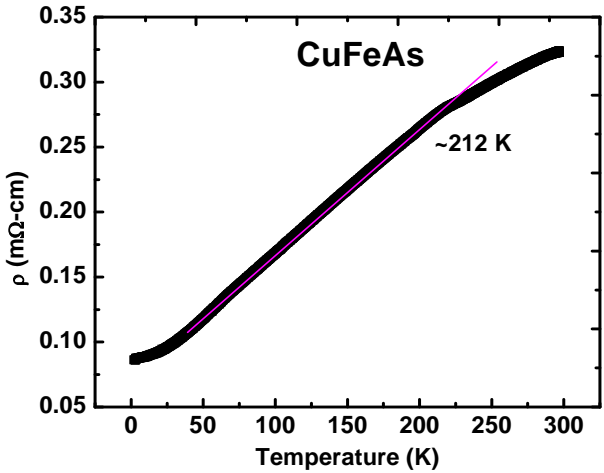

Fig. 5. Variable temperature resistivity plot for CuFeAs from 300 – 2 K. **(Color online)**

Keeping in mind that superconductivity in Fe-As materials is considered to be very much dependent on the structural features of the Fe-As layer, a larger difference in the tetrahedral bond angles ($\alpha$ and $\beta$) might be a reason for the absence of superconductivity in this material *vis-a-vis* LiFeAs. The anion height from Fe plane ($Z_{anion}$) also plays an important role in stabilizing the magnetic ground state (AFM or FM). It has been pointed out by Moon et al[21] and further supported by Qian *et al*,[7] that a larger $Z_{anion}$ usually drives the system towards ferromagnetic order and a low anion height render the system in an AFM ground state.[7,21] This possibly accounts for the fact that CuFeAs is antiferromagnetic and CuFeSb is ferromagnetic.[7] Li *et al*,[22] calculated the magnetic properties of LiFeAs using density functional theory method and showed theoretically that LiFeAs should have almost similar stripe-like AFM ground state as the other parent 1111 Fe-As based materials and FeSe. Due to tiny inter layer Fe-Fe interaction (*c*-parameter dependent) this magnetic ground state is not realized. Instead, one observes superconductivity. This implies that CuFeAs with smaller *c*-parameter and hence somewhat stronger Fe-Fe layer interaction than in LiFeAs realizes an AFM state, $T_N \sim 9$ K. We believe it may be possible to induce superconductivity in CuFeAs with the help of suitable doping (electron doping). We have plans to investigate these and other related phenomena in CuFeAs.

An important aspect of CuFeAs is that there is no crystallographic disorder among Cu and Fe ions just as is the case in CuFeSb. This opens up a possibility of trying monovalent nonmagnetic Cu-ions as dopant ions. This would lead to whole new approach to doping Fe-As materials RFeAsO, $MFe_2As_2$ and MFeAsF (M = Ca, Sr, Ba, Eu). This has not been tried so far. Our own efforts in this direction are in progress.

## 4. Conclusions

A new ternary Fe-As material CuFeAs has been synthesized. CuFeAs has the same structure as LiFeAs. It is unique among all the known members of the 111-family of Fe-pnictide materials in that it undergoes an antiferromagnetic transition below $T_N \sim 9$ K just as several undoped 1111-type RFeAsO do[††]. It would be extremely interesting to determine magnetic structure below its $T_N$ and also to study the lattice distortion associated with the magnetic transition as observed in LaFeAsO, for example.

## Acknowledgements

Authors thank Prof S. Patnaik for low temperature resistivity measurements and DST for providing the National SQUID facility at the department of physics, IIT Delhi. AKG thanks DST for providing financial support. GST and ZH thank CSIR and UGC, respectively, for a fellowship.

**Notes and references**

*ashok@chemistry.iitd.ac.in*

[†] Visting scientist at Solid State and Nanomaterials Research Laboratory, Department of Chemistry, IIT Delhi, India.

[††] NaFeAs exhibits both an antiferromagnetic transition ($T_N \sim 42$ K) and a superconducting transition ($T_c \sim 9$ K).